\documentstyle[12pt]{article}
\newcommand{\doublespacing}{\let\CS=\@currsize\renewcommand{\baselinesstrech}
{2.0}\tiny\CS}

\begin{document}

\title{New nonlinear coherent states and some of their nonclassical properties}
\author{B.Roy.\thanks{E-mail : barnana@isical.ac.in} and
P. Roy\thanks{E-mail : pinaki@isical.ac.in}\\ Physics \& Applied
Mathematics Unit\\ Indian Statistical Institute \\ Calcutta  700035\\ India} 


\maketitle

\vspace*{1.5cm}

\centerline{\bf Abstract}

\vspace{0.3cm}

\thispagestyle{empty}

\setlength{\baselineskip}{18.5pt}

We construct a displacement operator type nonlinear coherent
state and examine some of its properties. In particular it is shown that
this nonlinear coherent state exhibits nonclassical properties like
squeezing and sub-Poissonian behaviour.\\ 
PACS No: 42.50Dv, 42.50p
\newpage
\section{Introduction}
The importance of coherent states of various Lie algebras in different
branches of physics(in particular quantum optics) hardly needs to be
emphasized. For example the standard coherent state of the harmonic oscillator
corresponds to the Heisenberg-Weyl algebra. Similarly coherent states 
corresponding to Lie algebras like Su(1,1),Su(2) can also be constructed and they
have also found numerous
applications in quantum optics \cite{Pere,Zhang}. In this connection it may be
mentioned that coherent states are usually constructed using any of the following
three procedures : (1) Displacement operator technique (2) Annihilation operator
eigenstates (3) Minimum uncertainty states. However,these three approaches
are generally nonequivalent and only in the case of standard harmonic oscillator
coherent states obtained using any of the three approaches are equivalent.

On the other hand nonlinear coherent states \cite{Filho} or the f-coherent
states \cite{manko} 
are coherent states corresponding to nonlinear algebras rather than Lie
algebras. Nonlinear algebras are distinct from Lie algebras and have recently
been used to analyse a number of quantum mechanical systems \cite{Gran,vin,
JunRoy,kar}.
However nonlinear coherent states are not merely mathematical objects
but they are useful too. Recently it has been shown that 
nonlinear coherent states are useful in the description of the motion of a
trapped ion and various nonclassical properties of such states have also
been studied \cite{Filho}. We note that in refs \cite{Filho} and \cite{manko}
nonlinear coherent states have been defined as the right eigenstate of a generalised
annihilation operator $A$ (which emerges from the Hamiltonian describing the
dynamics). This is because in the case of nonlinear algebras
the commutator $[A,A^\dagger]$ is not a constant
or a linear function of the generators of the algebra but nonlinear in the
generators. As a consequence
it is difficult to obtain an explicit form of nonlinear coherent state 
constructed via the displacement operator technique.

In the present paper our aim is to construct nonlinear coherent states using
an operator which is similar to the displacement operator. This approach has
been used previously to construct nonlinear coherent states in the context
of isospectral Hamiltonians \cite{Fer1,Fer2} and deformed algebras \cite
{Bam,Jagan}. Subsequently we shall examine various nonclassical properties
like quadrature as well as amplitude squared squeezing,sub-Poissonian behaviour etc. of
the nonlinear coherent states so obtained.
The organisation
of the paper is as follows: in section 2 we shall describe the construction
of nonlinear coherent states; in section 3 we shall shall study nonclassical
properties of the nonlinear coherent state; finally section 4 is devoted to a
conclusion.

\section{Construction of nonlinear coherent states using a displacement type
operator}  

To begin with we note that the generalised annihilation (creation) operator
associated with nonlinear coherent states are given by 
\begin{equation}
A = a f(N)~~,~~A^\dagger = f(N) a^\dagger ~~,~~N = a^\dagger a \label{A}
\end{equation}
where $f(x)$ is a reasonably well behaved real function and $a^\dagger(a)$ is the
harmonic oscillator creation(annihilation) operator. It can be easily verified
that $A^\dagger,A$ and $N$ satisfy the following nonlinear algebra:
\begin{equation}
[N,A] = -A~~,~~[N,A^\dagger] = A^\dagger~~,~~[A , A^\dagger] = (N+1)f^2(N+1) - Nf^2(N) \label{comm}
\end{equation}
Clearly the nature of the nonlinear algebra depends on the choice of the
nonlinearity function $f(N)$. If however $f(N)=1$ then the nonlinear algebra 
in (\ref{comm}) reduces to the Heisenberg algebra.  

Nonlinear coherent states $|\alpha>$ are then defined as right eigenstates of
the generalised annihilation operator $A$ \cite{Filho,manko}:
\begin{equation}
A |\alpha> = \alpha |\alpha> \label{def}
\end{equation}
where $\alpha$ is an arbitrary complex number.
From (\ref{def}) we can now obtain an explicit form of the nonlinear coherent
state in a number state representation:
\begin{equation}
|\alpha> = C \sum_{n=0}^\infty \alpha^n d_n |n> \label{old}
\end{equation}
where the coefficients $d_n$'s are given by
\begin{equation}
d_0 = 1~,~d_n = [\sqrt n! f(n)!]^{-1}~,~f(n)! \equiv f(1)...f(n)
\end{equation}
and the normalisation constant C can be obtained from the condition
$<\alpha|\alpha> = 1$ and is given by 
\begin{equation}
C^2 = [\sum_{n = 0}^{\infty} d_n^2 |\alpha|^{2n}]^{-1}
\end{equation}

We now turn to the construction of a new type of nonlinear coherent state.
From the relation (\ref{comm}) we find that the r.h.s of the commutator
$[A,A^\dagger]$ is neither a constant nor linear in the generators but is a nonlinear
function of the generator $N$. As a result BCH disentangling theorem \cite{lou} can not
be applied and one can not use the displacement operator 
$exp(\alpha A-\alpha^*A^\dagger)$ to construct coherent states (see ref \cite{fuji}
for some recent results concerning the applicability of BCH disentangling theorem
to nonlinear algebras). 

We now proceed to determine an operator $B^\dagger$ 
which is conjugate of the operator $A$. In other words $A$ and $B^\dagger$ 
satisfy the commutation relation 
\begin{equation}
[A~,~B^\dagger] = 1 \label{newcom}
\end{equation}
while their hermitian conjugates $A^\dagger$ and $B$ satisfy the dual algebra
\begin{equation}
[B~,~A^\dagger] = 1 \label{newcom1}
\end{equation} 
Then from (\ref{A}),(\ref{newcom}) and (\ref{newcom1}) it follows that
\begin{equation}
B = a \frac{1}{f(N)}~~,~~B^\dagger = {1\over f(N)}a^\dagger  
\end{equation}
Let us now consider the following operators ($\beta$ being an arbitrary complex number):
\begin{equation}
\begin{array}{lcl}
D(\beta)& =& exp(\beta A^\dagger - \beta^* B)\\
D_1(\beta)& =& exp(\beta B^\dagger - \beta^* A)\\
\end{array}
\end{equation}
and note that for any two operators $X$ and $Y$ satisfying the relation
$[X,Y]=1$ the BCH theorem \cite{lou} yields
\begin{equation}
exp(\beta X - \beta^* Y) = exp(-\frac{\beta \beta^*}{2}) exp(\beta X) exp(-\beta^* Y) \label{bch}
\end{equation}
We now define nonlinear coherent states corresponding to the algebra
(\ref{newcom}) as $D(\alpha)|0>$ while those corresponding to the dual
algebra (\ref{newcom1}) as $D_1(\beta) |0>$.
Let us first consider the second case. Applying $D_1(\beta)$ on $|0>$ we
obtain on using (\ref{bch})
\begin{equation}
|\beta>_1 = D_1(\beta) |0> = c_1 \sum_{n=0}^\infty \frac{\beta^n}{\sqrt n!f(n)!} |n> \label{old1}
\end{equation}
where $c_1$ is a normalisation constant. Comparing (\ref{old}) and (\ref{old1})
it is seen that
the coherent state $|\beta>_1$ is the same
as the nonlinear coherent state $|\alpha>$. Thus the nonlinear coherent state
$|\alpha>$ can not only be obtained as an annihilation operator eigenstate
but it can also be obtained by the application of a displacement type 
operator. 

Let us now turn to the first possibility. Applying the operator $D(\beta)$ on
$|0>$ we get
\begin{equation}
|\beta> = D(\beta)|0> = c \sum_{n=0}^\infty \frac{\beta^n f(n)!}{\sqrt n!} |n> \label{new}
\end{equation}
where as before $c$ is a normalisation constant and we have used (\ref{bch})
to obtain (\ref{new}). The normalisation constant $c$ can be determined   
from the condition $<\beta|\beta> = 1$ and we get
\begin{equation}
c^2 = [\sum_{n=0}^\infty \frac{(\beta^* \beta)^n (f(n)!)^2}{n!}]^{-1}
\end{equation}
where $f(0)!\equiv 0$.
The superposition state obtained in (\ref{new}) is the new nonlinear
coherent state and this is distinct from
the nonlinear coherent state defined in (\ref{old})(provided of course we
use the same nonlinearity function $f(n)$ in both the cases).
In the next section we shall study various
properties of the nonlinear coherent state (\ref{new}).

\section{Non classical properties of nonlinear coherent state $|\beta>$}

In this section we shall examine squeezing and antibunching properties of
the new nonlinear coherent state $|\beta>$.
However before we proceed any
further it is necessary to specify the nonlinearity function $f(n)$. It is
clear that for every choice of $f(n)$ we shall have a different nonlinear
coherent states. In the present case we choose the following nonlinearity
function which is useful in the description of the motion of a trapped ion
\cite{Filho}:
\begin{equation}
f(n) = L_n^1(\eta^2)[(n+1)L_n^0(\eta^2)]^{-1} 
\end{equation}
where $L_n^m(x)$ are generalised Laguerre polynomials and $\eta$ is known
as the Lamb-Dicke parameter. Clearly $f(n)=1$ when $\eta=0$ and in this case
nonlinear coherent states become the standard coherent states. However when 
$\eta \neq 0$ nonlinearity starts developing with the degree of nonlinearity
depending on the magnitude of $\eta$ \cite{Filho,Roy}.

{\bf 3.1 Quadrature Squeezing}

Here we shall study quadrature squeezing of the 
new nonlinear coherent state $|\beta>$. 
In order to
do so let us consider the following hermitian quadrature operators:
\begin{equation}
X_1 = {(a + a^\dagger)\over 2}~~~,~~~Y_1 = {(a - a^\dagger)\over 2i} \label{qdef}  
\end{equation}
Then $X_1$ and $Y_1$ satisfy the following uncertainty relation:
\begin{equation}
<\Delta X_1^2>~~<\Delta Y_1^2>~~\geq~~{1\over 16}  \label{unc}
\end{equation}
where $<\Delta X^2> = <X^2> - <X>^2$.
From (\ref{unc}) it follows that a state is squeezed if any of the following
conditions hold:
\begin{equation}
\begin{array}{lcl}
<\Delta X_1^2 >~~& < &~~{1\over 4} \\ 
             \\ or \\
<\Delta Y_1^2 >~~& < &~~{1\over 4} \\  \label{ineq}
\end{array}
\end{equation}  
Now using (\ref{new}) and (\ref{qdef}) the squeezing conditions in (\ref{ineq})
can be reduced to the following forms:
\begin{equation}
\begin{array}{lcl}
F_1 &=& <a^{\dagger^2}> + <a^\dagger a> - 2<a^\dagger>^2 ~=~ \beta^2 I_2 + I_3 - 2\beta^2 I_1^2 < 0\\
                   \\or\\
G_1 &=& <a^\dagger a> -<a^\dagger>^2 ~=~I_3 -\beta^2 I_2 < 0\\ \label{ineq1}
\end{array}
\end{equation}
where $\beta$ is taken to be real and $I_i, i=1,2,3$ are infinite series 
whose explicit forms are given in the appendix.
 
We now evaluate the inequalities in (\ref{ineq1}) and the results are
presented in fig 1. In fig 1 we have plotted graphs of $F_1$ 
and $G_1$ against $\beta$ for fixed $\eta$. From fig 1 it is seen that while 
the curve of
$F_1$ is greater than zero that of $G_1$ is less than zero for a wide range of
$\beta$. Thus one of the inequalities in (\ref{ineq1}) is satisfied. 
This implies that the nonlinear coherent
state exhibits quadrature squeezing. We would like to mention here that we
have examined the inequalities in (\ref{ineq}) for a wide range of 
values of $\beta$ and $\eta$ and obtained the same qualitative behaviour
as in fig 1.

{\bf 3.2 Amplitude squared squeezing}

In order to examine whether or not the nonlinear coherent state exhibits amplitude
squared squeezing we introduce the following hermitian operators:
\begin{equation}
X_2 = {(a^2 + a^{\dagger ^2})\over 2}~~,~~Y_2 = {(a^2 - a^{\dagger ^2})\over 2i} \label{adef}
\end{equation}
Then $X_2$ and $Y_2$ obey the uncertainty relation
\begin{equation}
<\Delta X_2^2>~~<\Delta Y_2^2>~~\geq~~{1\over 4} |<[X_2,Y_2]>|^2 \label{qineq}
\end{equation}
From (\ref{qineq}) it follows that the nonlinear coherent state will exhibit
amplitude squared squeezing if
\begin{equation}
\begin{array}{lcl}
<\Delta X_2^2>~~& < &~~{1\over 2} |<[X_2,Y_2]>| \\
              \\ or \\
<\Delta Y_2^2>~~& < &~~{1\over 2} |<[X_2,Y_2]>| \label{ascond}
\end{array}
\end{equation}
Now proceeding as before the conditions (\ref{ascond}) for amplitude squared
squeezing become
\begin{equation}
\begin{array}{lcl}
F_2~&=&~<a^{\dagger^4}> + <a^{\dagger^2}a^2> - <a^{\dagger^2}>~=~\beta^4 I_4 + I_5 - I_2^2~~ < ~~0 \\
                \\or\\
G_2~&=&~<a^{\dagger^4}> -<a^{\dagger^2}a^2>~=~I_5 - \beta^4 I_4~~<~~0 \label{ascond1}
\end{array}
\end{equation}
where $I_4$ and $I_5$ are infinite series whose exact forms are given in the
appendix.
To examine the inequalities in (\ref{ascond1}) we plot $F_2$ and $G_2$ against
$\beta$ for $\eta$ fixed.
From figure 2 we find that $F_2 < 0$ for a certain range of $\beta$ and subsequently
it becomes positive. On the other hand $G_2$ is always positive. This implies
that the nonlinear coherent state $|\beta>$ exhibits amplitude squared
squeezing in the $X_2$ component. As in the case of quadrature squeezing we
have examined the inequalities in (\ref{ascond1}) for different values of 
$\beta$ and $\eta$ and it has been found that although squeezing can be
increased (or decreased) by changing the parameter values the basic qualitative
features remain the same as in fig 2.

{\bf 3.3 Sub-Poissonian Behaviour}

To examine sub-Poissonian behaviour we consider the second order correlation
function $g^2(0)$ defined by
\begin{equation}
g^2(0) = {<a^{\dagger^2}a^2> \over <a^\dagger a>^2} = {I_4\over I_3^2} \label{bunch}
\end{equation}
Then the state exhibits super-Poissonian/Poissonian/sub-Poissonian behaviour
according to
\begin{equation}
g^2(0) \begin{array}{c} >\\ =\\ <\\ \end{array}1  \label{bunch1}
\end{equation}
We now plot $g^2(0)$ against $\beta$ keeping $\eta$ constant. From fig 3 it
is seen that $g^2(0) < 1$ for the range of $\beta$ considered. This implies
that the nonlinear coherent state $|\beta>$ exhibits sub-Poissonian behaviour.
It may be noted that $g^(2)(0)$ has an increasing trend and so for sufficiently
large values of $\beta$ it may show super-Poissonian behaviour.

\section{Conclusion}

In this paper we have used a displacement type operator to construct a new
class of nonlinear coherent states which are distinct from those which are
annihilation operator eigenstates \cite{Filho,manko}. It has been shown that this nonlinear
coherent state exhibits interesting nonclassical properties like squeezing and
sub-Poissonian behaviour. We feel it would be interesting to examine other properties
e.g.,quantum interference ,phase properties etc of the nonlinear coherent
state $|\beta>$.

\centerline {\bf Appendix}
In order to examine squeezing and antibunching we need to evaluate several
expectation values like $<a^\dagger>,<a^{\dagger^2}>,<a^\dagger a>$ etc. These
expectation values are given in terms of the following series:
$$ \beta^{-1}<a> = \beta^{*^{-1}}<a^\dagger> = I_1 = c^2 \sum_{n=0}^\infty{(\beta^*\beta)^n f(n)! f(n+1)!\over n!} \eqno {(A1)}$$ 
$$ \beta^{-2}<a^2> = \beta^{*^{-2}}<a^{\dagger^2}> = I_2 = c^2 \sum_{n=0}^\infty{(\beta^*\beta)^n f(n)! f(n+2)!\over n!} \eqno {(A2)}$$ 
$$ <a^\dagger a> = I_3 = c^2 \sum_{n=0}^\infty{(\beta^*\beta)^{n+1} [f(n+1)!]^2\over n!} \eqno {(A3)}$$ 
$$ <a^{\dagger^2} a^2> = I_4 = c^2 \sum_{n=0}^\infty{(\beta^*\beta)^n [f(n+2)!]^2\over n!} \eqno {(A4)}$$ 
$$ \beta^{-4}<a^4> = \beta^{*^{-4}}<a^{\dagger^4}> = I_5 = c^2 \sum_{n=0}^\infty{(\beta^*\beta)^n f(n)! f(n+4)!\over n!} \eqno {(A5)}$$

\end{document}